\documentclass[pra,aps,twocolumn,superscriptaddress,showpacs]{revtex4}
\usepackage{amssymb}
\usepackage{amsfonts}
\usepackage{amsmath}
\usepackage{graphicx}
\usepackage{times}
\usepackage{graphics}
\usepackage{bm}
\usepackage{mathdots}
\usepackage{calc}
\usepackage{dcolumn}

\begin{document}
\title{Duality relations in a two-path interferometer with an asymmetric
beam splitter}
\author{Li Li}
\email{eidos@ustc.edu.cn}
\affiliation{Hefei National Laboratory for Physical Sciences at Microscale and Department
of Modern Physics, University of Science and Technology of China, Hefei,
Anhui 230026, China}
\author{Nai-Le Liu}
\email{nlliu@ustc.edu.cn}
\affiliation{Hefei National Laboratory for Physical Sciences at Microscale and Department
of Modern Physics, University of Science and Technology of China, Hefei,
Anhui 230026, China}
\author{Sixia Yu}
\affiliation{Hefei National Laboratory for Physical Sciences at Microscale and Department
of Modern Physics, University of Science and Technology of China, Hefei,
Anhui 230026, China}
\affiliation{Centre for Quantum Technologies, National University
of Singapore, 2 Science Drive 3, Singapore 117542}

\begin{abstract}
We investigate quantitatively the wave-particle duality in a general
Mach-Zehnder interferometer setup with an \textit{asymmetric} beam splitter.
The asymmetric beam splitter introduces additional \textit{a priori}
which-path knowledge, which is different for a particle detected at one
output port of the interferometer and a particle detected at the other.
Accordingly, the fringe visibilities of the interference patterns emerging
at the two output ports are also different. Hence, in sharp contrast with
the symmetric case, here we should concentrate on one output port and
distinguish two possible paths taken by the particles detected at that port
among four paths. It turns out that two nonorthogonal unsharp observables
are measured jointly in this setup. We apply the condition for joint
measurability of these unsharp observables to obtain a trade-off relation
between the fringe visibility of the interference pattern and the which-path
distinguishability.
\end{abstract}

\maketitle

\section{INTRODUCTION}

The wave-particle duality is a striking manifestation of Bohr's principle of
complementarity \cite{bohr} which lies at the heart of quantum mechanics. In
1979, Wootters and Zurek \cite{wootters} first \textit{quantified} the
wave-particle duality in an Einstein's version of the double-slit
experiment. Later, two kinds of \textit{duality inequalities} were
established in the standard Mach-Zehnder interferometer (MZI) setup. The
first one \cite{GY}, $\mathcal{P}^{2}+\mathcal{V}_{0}^{2}\leq 1,$ is about
the trade-off between the \textit{predictability} $\mathcal{P}$ of two
possible paths taken by a particle passing through the interferometer and
the \textit{a priori} fringe visibility $\mathcal{V}_{0}$ of the
interference pattern emerging at one output port of the interferometer. This
can be tested when the probabilities of taking the two paths are not equal
so that $\mathcal{P}\neq 0$. The second one \cite{jeager,englert,englert1},
e.g.,
\begin{equation}
\mathcal{D}^{2}+\frac{1-\mathcal{P}^{2}}{\mathcal{V}_{0}^{2}}\mathcal{V}%
^{2}\leq 1,  \label{di}
\end{equation}%
is about the trade-off between the \textit{distinguishability} $\mathcal{D}$
of the paths and the fringe visibility $\mathcal{V}$ when each particle is
coupled to another physical system which serves as a which-path detector
(WPD).

Another celebrated quintessential feature of quantum mechanics is that there
exist incompatible observables, i.e., observables which cannot be jointly
measured in a single device. However, in some cases, two incompatible sharp
observables could still be jointly measured on condition that some
imprecision is allowed. Exactly speaking, the unsharp versions of these
observables could be marginals of a bivariate joint observable, so that
measuring the joint observable offers simultaneously the values of the two
unsharp observables \cite{muynck1,muynck2}. The so-called joint
measurability problem---given two unsharp observables, are they jointly
measurable?---was first brought forward by Busch \cite{busch1} who solved it
in a very special case. Though there have been many partial results
concerning this problem in the past few years \cite%
{busch2,ueda1,ueda2,busch3,stano1,us}, the necessary and sufficient
condition for joint measurability of two general unsharp observables of a
two-level system was derived only recently by three independent groups \cite%
{stano2,jm,busch4}.

The problem of joint measurability has also been studied from other aspects,
including the uncertainty relation \cite{Andersson,Brougham}, quantum
cloning \cite{Brougham1,Ferraro}, and Bell inequalities \cite{Wolf}, and so
on. Recently, we have brought to light an intimate relationship between
joint measurability of two unsharp qubit observables and the wave-particle
duality illustrated in the standard MZI \cite{us}. In fact, the measurement
made on the WPD provides us the which-path information, or the
\textquotedblleft likelihood for guessing the right path\textquotedblright\ $%
\mathcal{L}=(1+\mathcal{D})/2,$ and meanwhile the counting detections at the
output ports of the interferometer yield the interference pattern. Since
these two measurements are made on different systems and therefore can be
made simultaneously, the whole setup provides \textit{de facto} a joint
measurement of two unsharp observables of the particle. Due to the fact that
the beam splitters in the standard MZI are \textit{symmetric}, i.e., the
proportion of the transmissivity and the reflectivity of each beam splitter
is $50:50$, the two unsharp observables turn out to be \textit{orthogonal}.
The condition for their joint measurability leads exactly to the duality
inequality Eq. (\ref{di}).

As well as we know, all duality inequalities so far have been derived in the
standard MZI setup with symmetric beam splitters. In the present work we
shall consider the wave-particle duality in a \textit{general} MZI setup
with \textit{asymmetric} beam splitters. When equipped with a WPD, this
general MZI setup provides a simultaneous measurement of two \textit{%
non-orthogonal} unsharp observables. The condition for joint measurability
of these two observables \cite{jm} enables us to obtain a duality
inequality. Unlike the symmetric case, two different interferences between
two paths among four paths appear at the two output ports in the asymmetric
case, and the \textit{a priori} path information is different for a particle
detected at one output port and a particle detected at the other, thus
needing to be treated more meticulously.

\section{GENERAL MZI Setup}

Consider the two-path MZI setup as depicted schematically in Fig. 1. For a
particle passing through the interferometer, the two distinct paths after
the first beam splitter $BS_{1}$ define two orthonormal states $\left\vert
0\right\rangle $ and $\left\vert 1\right\rangle $ which span a
two-dimensional Hilbert space. Without loss of generosity we can take $BS_{1}
$ as symmetric since the initial state of the particle is taken to be
arbitrary. The second beam splitter $BS_{2}$ is taken to be asymmetric and
we denote by $r$ its reflectivity, i.e., the probability of the particle
being reflected, and $t=1-r$ the transmissivity. The action of $BS_{2}$ on
the particle is effectively a unitary transformation%
\begin{equation}
B=\left(
\begin{array}{cc}
\sqrt{r} & \sqrt{t} \\
\sqrt{t} & -\sqrt{r}%
\end{array}%
\right)
\end{equation}%
on the above-mentioned two-dimensional Hilbert space.

\begin{figure}[h]
\includegraphics[width=85mm]{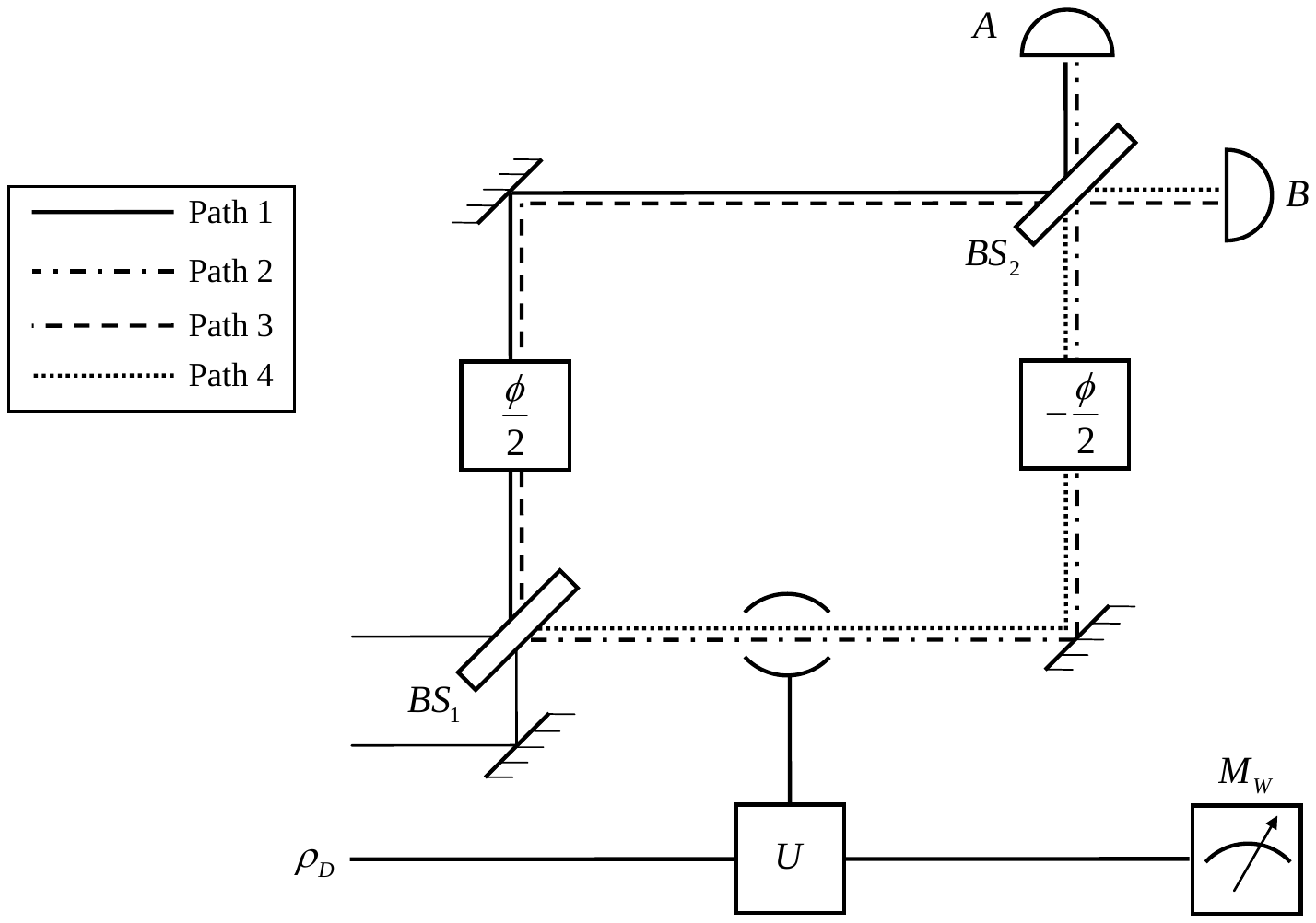}
\caption{The Mach-Zehnder interferometer with an asymmetric beam splitter.
The particle is prepared in an arbitrary state before entering the
interferometer and after the first beam splitter two different phases are
introduced. After the second beam splitter two counting detectors $A$ and $B$
record two interference patterns. The interference pattern recorded by the
detector $A$ (the detector $B$) comes from path 1 and path 2 (path 3 and
path 4). In addition, a WPD is introduced whose initial state is $\protect%
\rho _{D}$. After interacting with the particle via a controlled unitary
transformation, a special observable $W$ of the WPD is measured, from which
the path information of the particle is inferred.}
\end{figure}

To obtain simultaneously the which-path information and the interference
pattern, a WPD with initial state $\rho _{D}$ is coupled to the particle and
two phase shifts $\pm \phi $ are introduced for the two paths $|0\rangle
,|1\rangle $ respectively. The interaction between the particle and the WPD
is effectively a controlled unitary transformation $\left\vert
0\right\rangle \left\langle 0\right\vert \otimes I_{D}+\left\vert
1\right\rangle \left\langle 1\right\vert \otimes U$ where $I_{D}$ and $U$
are the identity and a unitary operator act on the Hilbert space of the WPD.
Thus the evolution of the particle and the detector after $BS_{1}$ is
governed by the unitary operator
\begin{equation}
U_{QD}=e^{i\phi /2}\left\vert \varphi _{0}\right\rangle \left\langle
0\right\vert \otimes I_{D}+e^{-i\phi /2}\left\vert \varphi _{1}\right\rangle
\left\langle 1\right\vert \otimes U
\end{equation}%
where $\left\vert \varphi _{0}\right\rangle \equiv B\left\vert
0\right\rangle $ and $\left\vert \varphi _{1}\right\rangle \equiv
B\left\vert 1\right\rangle $. Let $\rho $ be the state of the particle after
it has passed through $BS_{1}$. Then the final state of the whole system is
described by
\begin{eqnarray}
\rho _{f}^{(QD)} &=&U_{QD}(\rho \otimes \rho _{D})U_{QD}^{\dag }  \notag \\
&=&\sum_{a,b=0}^{1}e^{i(b-a)\phi }|\varphi _{a}\rangle \langle a|\rho
|b\rangle \langle \varphi _{b}|\otimes U^{a}\rho _{D}U^{\dagger b}.
\end{eqnarray}

\section{DISTINGUISHING TWO PATHS AMONG FOUR PATHS}

First of all, we notice that the asymmetric beam splitter $BS_{2}$ can be
regarded as a kind of which-path detector. Consider a simple case where the
probabilities for the particle taking the two paths after $BS_{1}$ are
equal, i.e., there is no \textit{a priori} which-path knowledge. If the
reflectivity $r$ of $BS_{2}$ is larger than $1/2$ and the particle is
detected at the detector $A$, immediately one can infer that the particle
passes more likely through the path $\left\vert 0\right\rangle $. So the
asymmetric beam splitter introduces additional \textit{a priori} which-path
knowledge and accordingly the visibility observed at the detector $A$ would
be decreased. In a general case where the probabilities for taking the two
paths after $BS_{1}$, namely $w_{+}=\left\langle 0\right\vert \rho
\left\vert 0\right\rangle $ and $w_{-}=\left\langle 1\right\vert \rho
\left\vert 1\right\rangle $, are not equal, the \textit{a priori} which-path
knowledge provided by $BS_{2}$ is different for a particle detected at one
output port of the interferometer and a particle detected at the other
output port. Accordingly, two different interference patterns emerge at the
two output ports. Hence, in order to explore which path the particle passes
through and to observe the interference pattern simultaneously, at first one
should concentrate on particles detected at one output port and then choose
a strategy to extract the which-path information from the measurements made
on the detector, taking into account the corresponding \textit{a priori}
which-path knowledge.

Specifically, the introduction of an asymmetric beam splitter entails the
need to distinguish \textit{two paths among four paths, }see Fig. 1. On the
observation of an interference pattern, e.g., at the detector $A$, two
possible paths taken by the particles contributing to the interference are
labeled with 1 and 2 respectively. When there is no WPD, the probabilities
of taking the two paths are
\begin{equation}
w_{1}=\frac{rw_{+}}{rw_{+}+tw_{-}},\quad w_{2}=\frac{tw_{-}}{rw_{+}+tw_{-}},
\end{equation}%
which provides the conditional \textit{a priori} which-path knowledge
provided that detector $A$ fires. Similarly, for the particles detected by $B
$, the probabilities of taking path 3 and path 4 are
\begin{equation}
w_{3}=\frac{tw_{+}}{rw_{-}+tw_{+}},\quad w_{4}=\frac{rw_{-}}{rw_{-}+tw_{+}}.
\end{equation}%
respectively. Note that (1) $w_{1,2}$ and $w_{3,4}$ are equal to $w_{\pm }$
when $BS_{2}$ is symmetric, and so to distinguish path 1 and path 2 (path 3
and path 4) is equivalent to distinguish the paths $\left\vert
0\right\rangle $ and $\left\vert 1\right\rangle $. Hence in a standard MZI
scheme it is not necessary to involve four paths; (2) when $w_{\pm }=1/2$,
the conditional \textit{a priori} path information $w_{1,2}$ is the same as $%
w_{3,4}$, but not equal to $w_{\pm }$. In fact, in the experiment scheme in
\cite{jacques} only two paths need to be distinguished and the difference
between $w_{\pm }$ and $w_{1,2}$ $(w_{3,4})$ is used to gain the which-path
information. In other words, the asymmetric beam splitter in the experiment
plays also the role of a WPD so that the path information and the
interference pattern are obtained via the same detector. Thus the setup in
\cite{jacques} can be regarded as a standard MZI.

In the scheme we considered here, the predictability $\mathcal{P}$, the
\textit{a priori} fringe visibility $\mathcal{V}_{0}$, the visibility $%
\mathcal{V}$, and the distinguishability $\mathcal{D}$ must be defined
particularly for each pair of paths. In what follows we shall consider only
the interference pattern registered by detector $A$, and the case for the
detector $B$ is similar. In this case the predictability of the paths 1 and
2 is obviously%
\begin{equation}
\mathcal{P=}\left\vert w_{1}-w_{2}\right\vert .
\end{equation}%
The two paths taken by the particle are conditioned on the clicks in
detector $A$, whose probability is given by
\begin{eqnarray*}
p\left( \phi \right)  &=&\text{\textrm{tr}}_{QD}\left[ |0\rangle \langle
0|\rho _{f}^{(QD)}\right] =rw_{+}+tw_{-} \\
&&+2\sqrt{rt}\left\vert \left\langle 0\right\vert \rho \left\vert
1\right\rangle \text{\textrm{tr}}_{D}(\rho _{D}U)\right\vert \cos \left(
\phi +\alpha +\delta \right) ,
\end{eqnarray*}%
where $\left\langle 0\right\vert \rho \left\vert 1\right\rangle =\left\vert
\left\langle 0\right\vert \rho \left\vert 1\right\rangle \right\vert
e^{i\alpha }$ and \textrm{tr}$_{D}(\rho _{D}U)=|$\textrm{tr}$_{D}(\rho
_{D}U)|e^{-i\delta }$. If the WPD is not turned on, i.e, $U=I_{D}$, then the
\textit{a priori} visibility reads
\begin{equation}
\mathcal{V}_{0}=\frac{p_{\max }-p_{\min }}{p_{\max }+p_{\min }}=\frac{2\sqrt{%
rt}\left\vert \left\langle 0\right\vert \rho \left\vert 1\right\rangle
\right\vert }{rw_{+}+tw_{-}}.
\end{equation}%
It is easy to see that the well-known duality relation \cite{GY}%
\begin{equation}
\mathcal{P}^{2}+\mathcal{V}_{0}^{2}\leq 1
\end{equation}%
still holds in the present case. If the WPD is turned on, then the fringe
visibility reads%
\begin{equation}
\mathcal{V}=\mathcal{V}_{0}\left\vert \text{\textrm{tr}}_{D}\left( \rho
_{D}U\right) \right\vert .
\end{equation}

A general strategy $\mathcal{S}$ to guess the path taken by the particle is
to divide the outcomes $W$ of the measurement of an observable $\hat{W}$
performed on the WPD into two disjoint sets, $S$ and $\bar{S}$. If $W\in S$,
then one guesses the path to be 1; if $W\in \bar{S}$, then one guesses the
path to be 2. The probability of guessing the right path is given by \cite%
{confluence}
\begin{equation}
\mathcal{L}_{\hat{W},S}=w_{1}\sum_{W\in S}\left\langle W\right\vert \rho
_{D}\left\vert W\right\rangle +w_{2}\sum_{W\in \bar{S}}\left\langle
W\right\vert U\rho _{D}U^{\dag }\left\vert W\right\rangle .
\end{equation}%
This is because once the particle is detected in path 1 (with probability $%
w_{1}$), the WPD will be in the state $\rho _{D}$; once the particle is
detected in path 2 (with probability $w_{2}$), the WPD will be in the state $%
U\rho _{D}U^{\dag }$. Let us denote
\begin{equation}
\eta _{S}\equiv \sum_{W\in S}\left\langle W\right\vert \rho _{D}\left\vert
W\right\rangle ,\quad \eta _{S}^{U}\equiv \sum_{W\in S}\left\langle
W\right\vert U\rho _{D}U^{\dag }\left\vert W\right\rangle ,  \label{d}
\end{equation}%
together with $\eta _{\bar{S}}=1-\eta _{S}$ and $\eta _{\bar{S}}^{U}=1-\eta
_{S}^{U}$. The which-path \textit{distinguishability} for the given strategy
$\mathcal{S}$ is then%
\begin{equation}
{\mathcal{D}}_{S}=2\mathcal{L}_{\hat{W},S}-1=2w_{1}\eta _{S}+2w_{2}\eta _{{S}%
}^{U}-1.  \label{D}
\end{equation}

\section{DUALITY\ RELATION\ FROM\ JOINT\ MEASURABILITY}

The duality relation in an interferometer turns out to be intimately related
to the joint measurement of two unsharp observables \cite{us}. Generally,
for a two-level system an unsharp observable is nothing else than a
two-outcome positive-operator-valued measure (POVM). Two general unsharp
observables $\left\{ O_{\pm }\right\} $ and $\left\{ O_{\pm }^{\prime
}\right\} $ of a qubit take the form%
\begin{equation}
O_{\pm }=\frac{I\pm \left( xI+\boldsymbol{m}\cdot \boldsymbol{\sigma }%
\right) }{2},\quad O_{\pm }^{\prime }=\frac{I\pm \left( yI+\boldsymbol{n}%
\cdot \boldsymbol{\sigma }\right) }{2}  \label{o}
\end{equation}%
where $I$ is the identity operator acting on the particle, and $\boldsymbol{%
\sigma }$ is the Pauli operator. The non-negativity imposes $\left\vert
x\right\vert +\left\vert \boldsymbol{m}\right\vert \leq 1$ and so on. When $%
x=0$ and $\left\vert \boldsymbol{m}\right\vert =1$, $O_{\pm }$ are
projectors of eigenstates of a sharp observable $\boldsymbol{m}\cdot
\boldsymbol{\sigma }$. So, generally an unsharp observable is the smeared
version of a sharp observable. The above two unsharp observables are jointly
measurable if and only if there exists a joint unsharp observable $\{M_{\mu
\nu }\}$ whose outcomes can be so grouped that the marginals correspond
exactly to the two given unsharp observables, i.e.,
\begin{equation}
O_{\mu }=\sum_{\nu =\pm }M_{\mu \nu },\quad O_{\nu }^{\prime }=\sum_{\mu
=\pm }M_{\mu \nu }.
\end{equation}%
If $y=0,$ then the necessary and sufficient condition for their joint
measurability reads \cite{jm}
\begin{equation}
\sqrt{\left( 1+x\right) ^{2}-\left\vert \boldsymbol{m}\right\vert ^{2}}+%
\sqrt{\left( 1-x\right) ^{2}-\left\vert \boldsymbol{m}\right\vert ^{2}}\geq
\frac{2\left\vert \boldsymbol{m}\times \boldsymbol{n}\right\vert }{\sqrt{%
\left\vert \boldsymbol{m}\right\vert ^{2}-(\boldsymbol{m}\cdot \boldsymbol{n}%
)^{2}}}.  \label{cond}
\end{equation}

In our general MZI setup, the unsharp observable $\mathcal{N}%
=\{N_{0},N_{1}=I-N_{0}\}$ corresponding to the interference pattern
registered in detector $A$ is given by $p(\phi )=$\textrm{tr}$_{Q}(\rho
N_{0})$ for an arbitrary $\rho $. Thus we obtain
\begin{equation*}
N_{0}=\mathrm{tr}_{D}\left[ U_{QD}^{\dag }\left( \left\vert 0\right\rangle
\left\langle 0\right\vert \otimes I_{D}\right) U_{QD}\left( I\otimes \rho
_{D}\right) \right] =\frac{I+{\boldsymbol{n}}\cdot \mathbf{\sigma }}{2}
\end{equation*}%
with%
\begin{equation}
\boldsymbol{n}=\left[ 2\frac{\mathcal{V}}{\mathcal{V}_{0}}\sqrt{rt}\cos
\left( \phi +\delta \right) ,2\frac{\mathcal{V}}{\mathcal{V}_{0}}\sqrt{rt}%
\sin \left( \phi +\delta \right) ,2r-1\right] .
\end{equation}%
For a given strategy ${\mathcal{S}},$ the probability of finding the
detector in one of the eigenstates in $S$ is given by \textrm{tr}$%
_{Q}(M_{S}\rho )$ for an arbitrary $\rho $ where
\begin{equation*}
M_{S}=\sum_{W\in S}\text{\textrm{tr}}_{D}\left[ U_{QD}^{\dag }(I\otimes
|W\rangle \langle W|)U_{QD}\left( I\otimes \rho _{D}\right) \right] .
\end{equation*}%
Thus the unsharp observable corresponding to the observable $\hat{W}$ and
the strategy $\mathcal{S}$ is $\mathcal{M}=\{M_{S},I-M_{S}\}$ with
\begin{equation*}
M_{S}=\frac{1}{2}\left[ \left( \eta _{S}+\eta _{S}^{U}\right) I+\left( \eta
_{S}-\eta _{S}^{U}\right) \sigma _{z}\right] =\frac{I+xI+\boldsymbol{m}\cdot
{\boldsymbol{\sigma }}}{2},
\end{equation*}%
in which notations in Eq. (\ref{d}) have been used and
\begin{equation}
x=\eta _{S}+\eta _{S}^{U}-1,\quad \boldsymbol{m}=\left( 0,0,\eta _{S}-\eta
_{S}^{U}\right) .
\end{equation}%
It is clear that as long as $r\neq 1/2$ we have $\boldsymbol{n\cdot m}\neq 0$%
, i.e., the two unsharp observables measured jointly in the general MZI
setup are \textit{non-orthogonal}, in contrast with a standard MZI setup
\cite{us}. From the joint measurement condition Eq. (\ref{cond}) it follows
that
\begin{equation}
\sqrt{\eta _{S}\eta _{S}^{U}}+\sqrt{\eta _{\bar{S}}\eta _{\bar{S}}^{U}}\geq
\frac{\mathcal{V}}{\mathcal{V}_{0}}.
\end{equation}%
Taking into the definition Eq. (\ref{D}) and similar to the derivation in
\cite{us}, we obtain
\begin{equation}
\mathcal{D}_{\mathcal{S}}^{2}+\frac{1-\mathcal{P}^{2}}{\mathcal{V}_{0}^{2}}%
\mathcal{V}^{2}\leq 1-\gamma _{S}^{2},
\end{equation}%
where $\gamma _{S}=2\left\vert w_{1}\sqrt{\eta _{S}\eta _{\bar{S}}}-w_{2}%
\sqrt{\eta _{S}^{U}\eta _{\bar{S}}^{U}}\right\vert $. By maximizing over all
possible strategies we obtain a duality inequality in the same form as Eq. (%
\ref{di}).

\section{CONCLUSIONS AND\ DISCUSSIONS}

We have considered how to illustrate quantitatively the wave-particle
duality a general MZI scenario with an asymmetric beam splitter. The
asymmetric beam splitter\textit{\ }introduces additional \textit{a priori}
which-path knowledge which is different for a particle detected at one
output port of the interferometer and a particle detected at the other, and
consequently the fringe visibilities of the interference patterns at the two
output ports are also different. Therefore, we should concentrate on
particles detected at one output port and distinguish two possible paths
taken by the particles detected at that port among four paths, and so our
result is not a straight-forward extension of the duality inequality in a
standard MZI set up with symmetric beam splitter. For each particle detected
at the output port, a pair of unsharp observables are jointly measured. It
turns out that non-orthogonality of the two unsharp observables is caused by
the asymmetric beam splitter, which characterizes the general MZI setup. We
have employed the condition for joint measurability of the two unsharp
observables to obtain a duality inequality.

It would be interesting to ask what is the experimental setup to measure
jointly a pair of most general unsharp observables of a two-level system,
though the condition for their jointly measurability has been established
\cite{jm}. Reversely, whether does the most general jointly measurability
condition imply a \textquotedblleft thorough" complementarity relation with
realizable and observable effects not limited by the known duality
inequalities? These questions remain open for further research. The answers
may lead to a device-independent duality inequality.

\begin{acknowledgments}
This work was supported by the NNSF of China, the CAS, the National
Fundamental Research Program (under Grant No. 2011CB921300).
\end{acknowledgments}


\begin{thebibliography}{99}
\bibitem{confluence} To see which path dose the particle actully pass
through, in standard MZI other two detectors can be directly set on the two
paths $\left\vert 0\right\rangle $ and $\left\vert 1\right\rangle $. The
conflunce of the paths 1 and 2 after $BS_{2}$ makes this method inavailable.
However, in principle one can still accomplish it with distinguishing
another freedom of the particle. As an example, polarization is used in \cite%
{englert exp}.
\end{thebibliography}
\end{document}